\documentstyle[aps,prl,floats]{revtex}
\begin{document}
\draft

\title{Links between gravity and dynamics of quantum liquids}
\author{G.E. Volovik}
\address{ Helsinki University of Technology, Low Temperature
Laboratory, P.O. Box 2200, FIN-02015 HUT, Finland\\
L. D. Landau Institute for Theoretical Physics, RAS, Kosygin Str. 2,
 117940 Moscow, Russia }
\maketitle

\date{\today}


\begin{abstract}
We consider the Landau-Khalatnikov two-fluid hydrodynamics of
superfluid liquid \cite{Khalatnikov} as an effective theory, which provides
a self-consistent
analog of Einstein equations for gravity and matter.
\end{abstract}

\

\centerline{\bf IV International Conference "Cosmology. Relativistic
Astropohysics.
Cosmoparticle Physics".}

\centerline{(COSMION-99)}

\centerline{In the Honor of 80-th Birthday of Isaak M. Khalatnikov}

\

\section{Introduction. Physical vacuum as condensed matter.}

In a modern viewpoint the relativistic quantum field theory is an
effective theory
\cite{Weinberg}. It is an emergent phenomenon arising in the low energy
corner of the
physical fermionic vacuum -- the medium, whose nature remains unknown. Also
it is argued that
in the low energy corner the symmetry must be enhanced \cite{Chadha}: If we
neglect the
very low energy region of electroweak scale, where some symmetries are
spontaneously violated,
then above this scale one can expect that the lower the energy, the better
is the Lorentz
invariance and other symmetries of the physical laws. The same phenomena
occur in the
condensed matter systems. If the spontaneous symmetry breaking at very low
energy is
neglected or avoided, then in the limit of low energy the symmetry of
condensed matter
is really enhanced. Moreover, there is one special universality class of
Fermi systems, where
in the low energy corner there appear almost all the symmetries, which we
know today in high
energy physics: Lorentz invariance, gauge invariance, elements of general
covariance, etc
(superfluid $^3$He-A is a representative of this class  \cite{parallel}).
The chiral
fermions as well as gauge bosons and gravity field arise as fermionic and
bosonic collective
modes of such a system. The inhomogeneous states of the condensed matter
ground state --
vacuum -- induce nontrivial effective metrics of the space, where the free
quasiparticles
move along geodesics.  This conceptual similarity between condensed matter
and quantum vacuum
allows us to simulate many phenomena in high energy physics and cosmology,
including axial
anomaly, baryoproduction and magnetogenesis, event horizon and Hawking
radiation, rotating
vacuum, expansion of the Universe, etc., probing these phenomena in
ultra-low-temperature
superfluid helium, atomic Bose condensates and superconductors. Some of the
experiments have
been already conducted.

The quantum field theory, which we have now, is incomplete due to ultraviolet
diveregences at small scales, where the ``microscopic''   physics of
vacuum becomes important. Here the analogy between quantum vacuum and
condensed matter could
give an insight into the transPlanckian physics. As in condensed matter
system, one can
expect that some or all of the known symmetries in Nature will be lost when
the Planck energy
scale is approached.  The condensed matter analogue gives examples of the
physically imposed
deviations from Lorentz invariance. This is important in  many different
areas of high energy
physics and cosmology, including possible CPT violation and black holes,
where the infinite
red shift at the horizon opens the route to the transPlanckian physics.

The low-energy properties of different condensed matter substances
(magnets, superfluids,
crystals, superconductors, etc.)  are robust, i.e. they do not depend
much on the details of microscopic (atomic) structure of these substances.
The main role
is played by symmetry and topology of condensed matter: they determine the soft
(low-energy) hydrodynamic variables, the effective Lagrangian describing
the low-energy dynamics, and topological defects. The microscopic details
provide us only with the ``fundamental constants'', which enter the effective
phenomenological Lagrangian, such  as speed of ``light'' (say, the speed of
sound),
superfluid density, modulus of elasticity, magnetic susceptibility, etc.
Apart from these  fundamental constants, which can be rescaled, the systems
behave
similarly in the infrared limit if they belong to the same universality and
symmetry classes,
irrespective of their microscopic origin. The detailed information on the
system is lost in
such acoustic or hydrodynamic limit \cite{LaughlinPines}. From the
properties of the low
energy collective modes of the system -- acoustic waves in case of crystals
-- one cannot
reconstruct the atomic structure of the crystal since all the crystals have
similar acoustic
waves described by the same equations of the same effective theory, in a
given case the
classical theory of elasticity. The classical fields of collective modes
can be quantized to
obtain quanta of acoustic waves -- the phonons, but this quantum field
remains the effective
field which does not give a detailed information on the real quantum
structure of the
underlying crystal.

It is quite probable that in the same way the quantization of classical
gravity, which is one
of the infrared collective modes of quantum vacuum, will not add more to
our understanding of
the "microscopic" structure of the vacuum
\cite{Hu96,Padmanabhan,LaughlinPines}. Indeed,
according to this analogy,  such properties of our world, as gravitation,
gauge fields,
elementary chiral fermions, etc., all arise in the low energy corner  as a
low-energy soft
modes of the underlying ``condensed matter''. At high energy (of the Planck
scale) these soft
modes disappear: actually they merge with the continuum of the high-energy
degrees of freedom
of the ``Planck condensed matter'' and thus cannot be separated anymore
from the others. Since
the gravity appears as an effective field in the infrared limit, the only
output of its
quantization would be the quanta of the low-energy gravitational waves --
gravitons.

The main advantage of the condensed matter analogy is that in principle we know
the condensed matter structure at any relevant scale, including the interatomic
distance,  which corresponds to the Planck scale. Thus the condensed matter can
suggest possible routes from our present low-energy corner of
``phenomenology''  to the
``microscopic'' physics  at Planckian and trans-Planckian energies.

\section{Landau-Khalatnikov two-fluid model hydrodynamics as effective
theory of gravity.}

\subsection{Superfluid vacuum and quasiparticles.}

Here we consider the simplest effective field theory of superfluids, where
only the
gravitational field appears as an effective field. The case of the fermi
superfluids, where
also the gauge field and chiral fermions appear in the low-energy corner
together with
Lorentz invariance is discussed in \cite{parallel,LammiTalk}.

According to Landau and Khalatnikov \cite{Khalatnikov} a weakly excited
state of the
collection of interacting $^4$He atoms  can be considered as a small number
of elementary
excitations -- quasiparticles, phonons and rotons. In addition, the state
without excitation
-- the ground state or vacuum -- can have collective degrees of freedom.
The superfluid
vacuum can move without friction, and  inhomogeneity of the flow serves as
the gravitational
and/or other effective fields. The matter propagating  in the presence of
this background is
represented by fermionic (in Fermi superfluids) or bosonic (in Bose
superfluids)
quasiparticles, which form the so called normal component of the liquid.
Such two-fluid
hydrodynamics introduced by Landau and Khalatnikov
\cite{Khalatnikov} is the example of the effective field theory which
incorporates the motion
of both the superfluid background (gravitational field) and excitations
(matter). This is the
counterpart of the Einstein equations, which incorporate both gravity and
matter.

One must distinguish between the particles and quasiparticles in
superfluids. The
particles describes the system on a microscopic level, these are atoms of
the underlying
liquid ($^3$He or $^4$He atoms). The many-body system of the interacting
atoms form the
quantum vacuum -- the ground state. The conservation laws experienced by
the atoms and their
quantum coherence in the superfluid state determine the low frequency
dynamics -- the
hydrodynamics -- of the collective variables of the superfluid vacuum. The
quasiparticles --
fermionic and bosonic -- are the low energy excitations above the vacuum
state. They form
the normal component of the liquid which determines the thermal and kinetic
low-energy
properties of the liquid.

\subsection{Dynamics of superfluid vacuum.} \label{DynamicsSuperfluidVacuum}

In the simplest superfluid the coherent motion of the superfluid vacuum is
characterized
by two collective (hydrodynamic) variables: the particle density $n({\bf
r},t)$  of  atoms
comprising the liquid and superfluid velocity $ {\bf v}_{\rm s}({\bf r},t)$
of their
coherent motion.  In a strict microscopic theory $n=\sum_{\bf p}  n({\bf
p})$, where
$n({\bf p})$ is the particle distribution functions. The particle number
conservation
provides one of the equations of the effective theory of superfluids -- the
continuity
equation
\begin{equation}
{\partial n\over \partial t}+ \nabla\cdot{\bf J}=0~.
\label{ContinuityEquation}
\end{equation}
 The  conserved current of  atoms in monoatomic
superfluid lquid is
\begin{equation}
{\bf J}= \sum_{\bf p} {{\bf p} \over m} n({\bf p})~,
\label{TotalCurrent1}
\end{equation}
where  $m$   is  the bare mass of the particle. Note that  the liquids
considered here are
nonrelativistic and obeying the Galilean transformation law. In the
Galilean system the momentum of particles and the particle current
are related by Eq.(\ref{TotalCurrent1}). In the effective theory the
particle current
 has two contributions
\begin{equation}
{\bf J}= n {\bf v}_{\rm s}+
{\bf J}_{\rm q}~,~{\bf J}_{\rm q}=\sum_{\bf p} {{\bf p} \over m} f({\bf p})~.
\label{TotalCurrent2}
\end{equation}
 The  first term $n {\bf v}_{\rm s}$ is the current
transferred coherently by the collective motion of superfluid vacuum with
the superfluid
velocity ${\bf v}_{\rm s}$. If quasiparticles are excited above the ground
state, their
momenta also contribute to the particle current, the second term in rhs of
Eq.(\ref{TotalCurrent2}), where $f({\bf p})$ is the distribution function of
quasiparticles.    Note that under the Galilean transformation to the
coordinate system
moving with the velocity
${\bf u}$ the superfluid velocity transforms as ${\bf v}_{\rm s}\rightarrow
{\bf v}_{\rm s} +
{\bf u}$, while the momenta of particle and quasiparticle transform
differently:
${\bf p} \rightarrow {\bf p}  + m {\bf u}$ for microscopic particles and
${\bf p} \rightarrow
{\bf p}$ for quasiparticles.

The second equation for the collective variables is the London
equation for the superfluid velocity, which is curl-free in superfluid $^4$He
($\nabla\times{\bf v}_{\rm s}=0$):
\begin{equation}
m{\partial  {\bf v}_{(s)}\over \partial t}   +   \nabla {\delta {\cal
E}\over \delta
n} =0~.
\label{LondonEquation}
\end{equation}
Together with the kinetic equation for the quasiparticle   distribution
function $f({\bf p})$, the Eqs.(\ref{LondonEquation}) and
(\ref{ContinuityEquation}) for
collective fields ${\bf v}_{\rm s}$ and $n$ give the complete effective
theory for the
kinetics of quasiparticles (matter) and coherent motion of vacuum
(gravitational field) if
the energy functional
${\cal E}$ is known. In the limit of low density of quasiparticles, when
the interaction
between quasiparticles can be neglected, the simplest Ansatz satisfying the
Galilean
invariance is
\begin{equation}
{\cal E}=\int d^3r \left( {m\over 2}n{\bf v}_{\rm s}^2 + \epsilon(n) +
\sum_{\bf p} \tilde E({\bf p},{\bf r}) f({\bf p},{\bf r})\right)~.
\label{Energy}
\end{equation}
Here $\epsilon(n)$ is the (phenomenological) vacuum energy as a function of
the particle
density;
$\tilde E({\bf p},{\bf r})=E({\bf p},n({\bf r}))+ {\bf
p}\cdot{\bf v}_{\rm s}({\bf r})$ is the  Doppler shifted quasiparticle
energy in the
laboratory frame; $E({\bf p},n({\bf r}))$ is the quasiparticle energy
measured in the
frame comoving with the superfluid vacuum. The Eqs.
(\ref{ContinuityEquation}) and
(\ref{LondonEquation}) can be also obtained from the Hamiltonian formalism
using
Eq.(\ref{Energy}) as Hamiltonian and Poisson brackets
\begin{equation}
 \left\{{\bf v}_{\rm s}({\bf r}_1),n({\bf r}_2)\right\}={1\over m}\nabla
\delta({\bf
r}_1-{\bf r}_2)~,~ \left\{n({\bf r}_1),n({\bf r}_2)\right\}= \left\{{\bf
v}_{\rm s}({\bf
r}_1),{\bf v}_{\rm s}({\bf r}_2)\right\}=0~.
\label{PoissonBrackets}
\end{equation}

Note that the Poisson brackets between components of superfluid velocity
are zero only for
curl-free superfluidity. In a general case it is
\begin{equation}
 \left\{ v_{{\rm s}i}({\bf r}_1),v_{{\rm s}j}({\bf r}_2)\right\}=-{1\over
mn}e_{ijk}
 (\nabla\times {\bf
v}_{\rm s})_k \delta( {\bf r}_1-{\bf r}_2)~.
\label{PoissonBracketsVelocity}
\end{equation}
In this case even at $T=0$, when the quasiparticles are absent, the
Hamiltonian description of
the hydrodynamics is only possible: There is no Lagrangian, which can be
expressed in terms of
the hydrodynamic variables ${\bf v}_{\rm s}$ and $n$. The absence of the
Lagrangian in many
condensed matter systems is one of the consequences of the reduction of the
degrees
of freedom in effective field theory, as compared with the fully
microscopic description. In
ferromagnets, for example, the number of the hydrodynamic variables is odd: 3
components of the magnetization vector ${\bf M}$. They thus cannot form the
canonical pairs of
conjugated variables. As a result one can use either the Hamiltonian
description or introduce
the effective action with the Wess-Zumino term, which contains an extra
coordinate $\tau$:
\begin{equation}
S_{\rm WZ}\propto \int d^3x~dt~d\tau ~ {\bf M}\cdot(\partial_t{\bf
M}\times\partial_\tau{\bf M})~.
\label{WessZumino}
\end{equation}

\subsection{Normal component -- ``matter''.}

 In a local thermal equilibrium the distribution of quasiparticles is
characterized by
local temperature $T$ and by local velocity ${\bf v}_{\rm n}$ called the
normal component
velocity
\begin{equation}
f_{\cal T}({\bf p})=\left(\exp { \tilde E({\bf p})- {\bf p}  {\bf v}_n\over
T} \pm 1\right)^{-1}~,
\label{Equilibrium}
\end{equation}
where the sign + is for the fermionic quasiparticles in Fermi superfluids
and the   sign -
is for the bosonic quasiparticles in Bose superfluids. Since $\tilde E({\bf
p})=E({\bf
p}) +{\bf p}\cdot {\bf v}_{\rm s}$, the equilibrium distribution is
determined by the
Galilean invariant quantity ${\bf v}_{\rm n} - {\bf v}_{\rm s}\equiv {\bf
w}$, which is the
normal component velocity measured in the frame comoving with superfluid
vacuum. It is called
the counterflow velocity. In the limit when the conterflow velocity ${\bf
v}_{\rm n}-{\bf
v}_{\rm s}$ is small, the quasiparticle
(``matter'') contribution to the particle current is proportional to the
counterflow
velocity:
\begin{equation}
 J_{{\rm q} i} =n_{{\rm n}ik}(v_{{\rm n}k}-v_{{\rm s}k})~,
\label{EquilibriumCurrent}
\end{equation}
where the tensor $n_{{\rm n}ik}$ is the so called density of the normal
component. In this
linear regime the total current can be represented as the sum of the
currents of the normal
and superfluid components
\begin{equation}
J_{i} =n_{{\rm s}ik}v_{{\rm s}k}+n_{{\rm n}ik}v_{{\rm n}k} ~,
\label{TotalCurrent2}
\end{equation}
where tensor $n_{{\rm s}ik}=n\delta_{ik}-n_{{\rm n}ik}$ is the so called
density of superfluid component.
In the isotropic
superfluids,
$^4$He and $^3$He-B, the normal component is isotropic tensor, $n_{{\rm
n}ik}=n_{{\rm
n}}\delta_{ik}$, while in anisotropic superfluid
$^3$He-A the normal component density is a
uniaxial tensor \cite{VollhardtWolfle}.  At
$T=0$ there the quasiparticles are frozen out and one has
$n_{{\rm n}ik}=0$ and $n_{{\rm s}ik}=n\delta_{ik}$.

\subsection{Quasiparticle spectrum and effective metric}

The structure of the quasiparticle spectrum in superfluid $^4$He becomes more and more
universal the lower the energy. In the low energy corner the spectrum o
f these
quasiparticles, phonons, can be obtained in the framework of the effective
theory. Note that
the effective theory is unable to describe the high-energy part of the
spectrum -- rotons,
which can be determined in a fully microscopic theory only.  On the
contrary, the spectrum of
phonons is linear, $E({\bf p},n)\rightarrow c(n)|{\bf p}|$, and only the
``fundamental
constant'' -- the speed of ``light''  $c(n)$ -- depends on the physics of
the higher energy
hierarchy rank. Phonons represent the quanta of the collective modes of the
superfluid vacuum, sound waves, with the speed of sound obeying
$c^2(n)=(n/m)(d^2\epsilon/dn^2)$. All other information on the microscopic
atomic nature of
the liquid is lost. Note that for the curl-free superfluids the sound waves
represent the only
``gravitational'' degree of freedom. The Lagrangian for these
``gravitational waves''
propagating above the smoothly varying background is obtained by
decomposition of the
superfluid velocity and density into the smooth and fluctuating parts:
 ${\bf v}_{\rm s}=
{\bf v}_{\rm s~~smooth} +\nabla \alpha$ \cite{unruh,vissersonic}.  The
Lagrangian for the
scalar field $\alpha$ is:
\begin{equation}
{\cal L}=  {m\over 2}n \left( (\nabla \alpha)^2- {1\over
c^2}\left(\dot\alpha    +
({\bf v}_{\rm s}\cdot\nabla)\alpha\right)^2\right)\equiv {1\over
2}\sqrt{-g}g^{\mu\nu}\partial_\mu\alpha \partial_\nu\alpha~.
\label{LagrangianSoundWaves}
\end{equation}
Thus in the low energy corner the Lagrangian for sound waves has an
enhanced symmetry --
the Lorentzian form, where the effective Riemann metric  experienced by the
sound wave, the so
called acoustic metric, is simulated by the smooth parts of the
hydrodynamic fields:
\begin{equation}
 g^{00}=-{1\over mnc} ~,~ g^{0i}=-{v_{\rm s}^i\over mnc} ~,~ g^{ij}=
{c^2\delta^{ij} -v_{\rm s}^i v_{\rm s}^j\over mnc} ~,
\label{ContravarianAcousticMetric}
\end{equation}
\begin{equation}
 g_{00}=-{mn\over  c}(c^2-{\bf v}_{\rm s}^2) ~,~
g_{0i}=-{mnv_{{\rm s} i}\over c} ~,~ g_{ij}= {mn\over  c}\delta_{ij} ~,~
\sqrt{-g}={m^2n^2\over c}~.
\label{CovarianAcousticMetric}
\end{equation}
Here and further ${\bf v}_{\rm s}$ and $n$ mean the smooth parts of the
velocity and density
fields.

The energy spectrum of sound wave quanta, phonons, which represent the
``gravitons''
in this effective gravity, is determined by
\begin{equation}
 g^{\mu\nu}p_\mu p_\nu=0~,~~{\rm or} ~~ (\tilde E-{\bf p}\cdot {\bf v}_{\rm
s})^2=c^2p^2  ~.
\label{PhononEnergySpectrum}
\end{equation}

\subsection{Effective quantum field and effective action}

The effective action Eq.(\ref{LagrangianSoundWaves}) for phonons formally
obeys the
general covariance, this is an example of how the enhanced symmetry arises
in the
low-energy corner. In addition,  in the classical limit of
Eq.(\ref{PhononEnergySpectrum})
corresponding to geometrical optics (in our case this is geometrical
acoustics) the
propagation of phonons is invariant under the conformal transformation of
metric,
$g^{\mu\nu}\rightarrow \Omega^2 g^{\mu\nu}$. This symmetry is lost at the
quantum level: the
Eq.(\ref{LagrangianSoundWaves}) is not invariant under general conformal
transformations, however the reduced symmetry is still there:
Eq.(\ref{LagrangianSoundWaves}) is invariant under scale transformations with
$\Omega={\rm Const}$.

In superfluid $^3$He-A the  other effective fields and new symmetries
appear in the low energy
corner, including also the effective $SU(2)$ gauge fields and gauge
invariance. The symmetry
of fermionic Lagrangian induces, after integration over the quasiparticles
degrees of freedom,
the corresponding symmetry of the effective action for the gauge fields.
Moreover, in
addition to superfluid velocity field  there are appear the other
gravitational degrees of
freedom with the spin-2 gravitons. However, as distinct from the effective
gauge fields,
whose effective action is very similar to that in particle physics, the
effective gravity
cannot reproduce in a full scale the Einstein theory: the effective action
for the metric is
contaminated by the noncovariant terms, which come from the
``transPlanckian'' physics
\cite{parallel}.  The origin of difficulties with effective gravity in
condensed matter is
probably the same as the source of the problems related to quantum gravity
and cosmological
constant.

The quantum quasiparticles interact with the classical collective fields
${\bf v}_{\rm s}$
and $n$, and with each other. In Fermi superfluid $^3$He the fermionic
quasiparticles
interact with many collective fields describing the multicomponent order
parameter and with
their quanta. That is why one obtains the interacting Fermi and Bose
quantum fields,
which are in many respect similar to that in particle physics.
However, this field theory can be applied to a lowest orders of the
perturbation theory
only.  The higher order diagrams are divergent and nonrenormalizable, which
simply means that
the effective theory is valid when only the low energy/momentum
quasiparticles are involved
even in their virtual states. This means that only those terms in the
effective action can be
derived by integration over the quasiparticle degrees of freedom, whose
integral are
concentrated solely in the low-energy region. For the other processes one
must go beyond the
effective field theory and consider the higher levels of description, such
as Fermi liquid
theory, or further the microscopic level of the underlying liquid with
atoms and their
interactions. In short, all the terms in effective action come from
microscopic ``Planck''
physics, but only some fraction of them can be derived within the effective
field theory.

In Bose supefluids the fermionic degrees of freedom are absent, that is why
the quantum field
theory there is too restrictive, but nevertheless it is useful to consider
it since it
provides the simplest example of the effective theory. On the other hand the
Landau-Khalatnikov scheme is rather universal and is easily extended to
superfluids with
more complicated order parameter and with fermionic degrees of freedom (see
the book \cite{VollhardtWolfle}).

\subsection{Vacuum energy and cosmological constant}\label{VacuumEnergyAnd}

The vacuum energy density $\epsilon(n)$ and the parameters which
characterize the quasparticle
energy spectrum cannot be determined by the effective theory: they are
provided solely by the
higher (microscopic) level of description. The microscopic calculations
show that at zero
pressure the vacuum energy per one atom of the liquid $^4$He is about
$\epsilon(n_0)/n_0
\sim -7$K
\cite{Woo}.  It is instructive to compare this microscopic result  with the
estimation of the
vacuum energy if we try to obtain it from the effective theory. In
effective theory
the vacuum energy is given by the zero point motion of phonons
\begin{equation}
\epsilon_{\rm eff} = (1/2)\sum_{E({\bf p})<\Theta} cp ={1\over 16\pi^2}
{\Theta^4\over \hbar ^3 c^3} ={1\over 16\pi^2}\sqrt{-g}
\left(g^{\mu\nu}\Theta_{\mu}
\Theta_{\nu}\right)^2  ~.
\label{VacuumEnergy}
\end{equation}
Here $c$ is the speed of sound; $\Theta \sim \hbar c/a$ is the
Debye characteristic temperature with $a$ being the interatomic space,
$\Theta$ plays the
part of the ``Planck'' cutoff energy scale;
$\Theta_{\mu}=(-\Theta,0,0,0)$.

We wrote the Eq.(\ref{VacuumEnergy}) in the form which is different from the
conventional cosmological term $\Lambda \sqrt{-g}$. This is to show that
both forms and the
other possible forms too have the similar drawbacks. The
Eq.(\ref{VacuumEnergy}) is conformal
invariant due to conformal invariance experienced by the quasiparticle
energy spectrum in
Eq.(\ref{PhononEnergySpectrum}) (actually, since this term does not depend
on derivatives,
the conformal invariance is equivalent to invariance under multiplication
of $g_{\mu\nu}$ by
constant factor). However, in Eq.(\ref{VacuumEnergy}) the general
covariance is violated by
the cutoff. On the contrary, the conventional cosmological term
$\Lambda \sqrt{-g}$ obeys the general covariance, but it is not invariant
under
transformation $g_{\mu\nu}\rightarrow \Omega^2g_{\mu\nu}$ with constant
$\Omega$.  Thus
both forms of the vacuum energy violate one or the other symmetry of the
low-energy effective
Lagrangian Eq.(\ref{LagrangianSoundWaves}) for phonons, which means that
the vacuum energy
cannot be determined exclusively within the low-energy domain.

Now on the magnitude of the vacuum energy. The Eq.(\ref{VacuumEnergy})
gives $\epsilon_{\rm
eff}(n_0)/n_0 \sim 10^{-2}\Theta \sim 10^{-1}$K. The magnitude of the
energy is much smaller
than the result obtained in the microscopic theory, but what is more
important the energy has
an opposite sign. This means again that the effective theory must be used
with great caution,
when one calculates those quantities, which crucially (non-logarithmically)
depend on the
``Planck'' energy scale. For them the higher level ``transPlanckian''
physics must be used
only. In a given case the many-body wave function of atoms of the
underlying quantum liquid
has been calculated to obtain the vacuum energy \cite{Woo}. The quantum
fluctuations of the
phonon degrees of freedom in Eq.(\ref{VacuumEnergy}) are already contained
in this
microscopic wave function. To add the energy of this zero point motion of
the effective field
to the microscopically calculated energy $\epsilon(n_0)$ would be the
double counting. Thus
the proper regularization of the vacuum energy in the effective field
theory must by equating
it to exact zero.

This conjecture is confirmed by consideration of the equilibrium conditions
for the liquid.
The equilibrium condition for the superfluid vacuum is $(d\epsilon/d
n)_{n_0}=0$. Close to
the equilibrium state one has
\begin{equation}
\epsilon(n)=\epsilon(n_0) +{1\over 2} {m c^2\over n_0} (n-n_0)^2~.
\label{VacuumCloseToEquil}
\end{equation}
From this equation it follows that the variation of the vacuum energy over
the metric
determinant must be zero in equilibrium:
$d\epsilon/dg|_{n_0}=(d\epsilon/d n)_{n_0}/(d g/d n)_{n_0}= 0$. This
apparently shows that the
vacuum energy can be neither of the form of Eq.(\ref{VacuumEnergy}) nor in
the form
$\Lambda\sqrt{-g}$. The metric dependence of the vacuum energy  consistent
with the
equilibrium condition and Eq.(\ref{VacuumCloseToEquil}) could be only of
the type
$A+B(g-g_0)^2$, so that the cosmological term in Einstein equation would be
$\propto
(g-g_0)g_{\mu\nu}$. This means that in equilibrium, i.e. at $g=g_0$, the
cosmological
term is zero and thus the equilibrium vacuum is not gravitating. In
relativistic theories
such dependence of the Lagrangian on
$g$ can occur in the models where the determinant of the metric is the
variable which is not transformed under coordinate transformations, i.e.
only the invariance
under coordinate transformations with unit determinant represents the
fundamental symmetry.

This probably has some relation to the problem of the cosmological constant
in Einstein
theory of gravity, where the estimation in Eq.(\ref{VacuumEnergy}) with $c$
being the speed
of light and $\Theta=E_{\rm Planck}$ gives the cosmological term by 100
orders of magnitude
higher than its upper experimental limit. The gravity is the low-frequency,
and actually the
classical output of all the quantum degrees of freedom of the ``Planck
condensed matter''.
So one should not quantize the gravity again, i.e. one should not use the
low energy
quantization for construction of the Feynman diagrams technique with
diagrams containing the
integration over high momenta. In particular the effective field theory is
not appropriate
for the calculation of the vacuum energy and thus of the cosmological
constant. Moreover,
one can argue that, whatever the real ``microscopic'' energy of the vacuum
is, the energy of
the equilibrium vacuum is not gravitating: The diverging energy of quantum
fluctuations of
the effective fields and thus the cosmological term must be regularized to
zero as we
discussed above, since these fluctuations are already contained in the
``microscopic wave
function'' of the vacuum.

This however does not exclude the Casimir effect, which appears if the
vacuum is not
homogeneous. The smooth deviations from the homogeneous equilibrium vacuum
are within the
low-energy domain: they can be successfully described by the effective
field theory, and their
energy can gravitate.

\subsection{Einstein action and higher derivative
terms}\label{EinsteinActionAnd}

In principle, there are the nonhydrodynamic terms in the effective action,
which are not
written in Eq.(\ref{Energy}) since they contain space and time  derivatives
of the
hydrodynamics variable,
$n$ and  ${\bf v}_{\rm s}$,  and  thus are relatively small. Only part of
them can be
obtained using the effective theory. As in the case of Sakharov effective
gravity
\cite{Sakharov}, the standard integration over the massless scalar field
$\alpha$
propagating in inhomogeneous $n$ and  ${\bf v}_{\rm s}$ fields, which
provide the effective
metric, gives the curvature term in Einstein action. It can also be written
in two ways.
The form which respects the general covariance of the phononic Lagrangian
for $\alpha$ field
in Eq.(\ref{LagrangianSoundWaves}) is:
\begin{equation}
{\cal L}_{\rm  Einstein}=-{1 \over 16\pi G }
\sqrt{-g} R    ~~,
\label{EinsteinAction}
\end{equation}
This form does not obey the invariance under multiplication of $g_{\mu\nu}$
by constant
factor, which shows its dependence on the  ``Planck'' physics. The
gravitational Newton constant $G$  is expressed in terms of the ``Planck''
cutoff:
$G^{-1}\sim \Theta^2$. Another form, which explicitly contains the
``Planck'' cutoff,
\begin{equation}
{\cal L}_{\rm  Einstein}=-{1 \over 16\pi }
\sqrt{-g} R g^{\mu\nu}\Theta_{\mu}\Theta_{\nu}   ~~,
\label{EinsteinActionModified}
\end{equation}
is equally bad: the action is invariant under the scale transformation of
the metric, but the
general covariance is violated. Such incompatibility of different low-energy
symmetries is the hallmark of the effective theories.

To give an impression on the relative magnitude of the Einstein action
let us express the Ricci scalar in terms of the superfluid velocity field only
\begin{equation}
\sqrt{-g} R  ={1\over c^3}\left(2\partial_t \nabla\cdot  {\bf v}_{\rm s} +
\nabla^2(v_{\rm
s}^2)\right) ~.
\label{RicciScalar}
\end{equation}
In superfluids the Einstein action is small compared to the dominating kinetic
energy term $mn{\bf v}_{\rm s}^2/2$ in Eq.(\ref{Energy})   by   factor
$a^2/l^2$, where $a$ is again the atomic (``Planck'') length scale and $l$
is the
characteristic macroscopic length at which the velocity field changes. That
is why it can be
neglected in the hydrodynamic limit, $a/l \rightarrow 0$. Moreover, there
are many  terms of
the same order in effective actions which do not display the general
covariance, such as
 $(\nabla\cdot  {\bf v}_{\rm s})^2$. They are provided by microscopic
physics, and there is
no rule in superfluids according to which these noncovariant terms  must be
smaller
than the Eq.(\ref{EinsteinAction}).  But in principle, if the gravity field as
collective field arises from the other degrees of freedom, different from
the condensate
motion, the Einstein action can be dominating.

There are the higher order derivative terms, which are quadratic in the
Riemann tensor,
such as
\begin{equation}
  \sqrt{-g}R^2 ~\ln \left( {g^{\mu\nu}\Theta_{\mu}\Theta_{\nu}\over R}\right)~.
\label{SquareRicciScalar}
\end{equation}
They only logarithmically depend on the cut-off and thus their calculation
in the framework of
the effective theory is possible.  Because of the logarithmic divergence
(they are of
the relative order $(a/l)^4~\ln (l/a)$) these terms dominate over the
noncovariant terms of
order $(a/l)^4$,  which are obtained in fully microscopic calculations.
Being determined
essentially by the phononic Lagrangian in Eq.(\ref{LagrangianSoundWaves}),
these terms
respect (with logarithmic accuracy) all the symmetries of this Lagrangian
including the
general covariance and the invariance under rescaling the metric. That is
why they are the
most appropriate terms for the self-consistent effective theory of gravity.
The logarithmic
terms also appear in the effective action for the effective gauge fields,
which take place in
superfluid $^3$He-A \cite{LammiTalk}. These terms in superfluid $^3$He-A
have been obtained
first in microscopic calculations, however it appeared that their physics
can be completely
determined by the low energy tail and thus they can be calculated using the
effective theory.
This is well known in particle physics as running coupling constants and
zero charge effect.

Unfortunately in effective gravity of superfluids these logarithmic terms
are small compared
with the main terms -- the vacuum energy and the kinetic energy of the
vacuum flow. This means
that the superfluid liquid is not the best condensed matter for simulation
of Einstein
gravity. In $^3$He-A there are other components of the order parameter,
which also give rise
to the effective gravity, but superfluidity of $^3$He-A remains to be an
obstacle. One must
try to construct the non-superfluid condensed matter system which belongs
to the same
universality class as $^3$He-A, and thus contains the  effective Einstein
gravity as emergent
phenomenon, which is not contaminated by the superfluidity.

\section{``Relativistic'' energy-momentum tensor for ``matter'' moving in
``gravitational'' superfluid background in two fluid hydrodynamics}

\subsection{Kinetic equation for quasiparticles (matter)}

The distribution function $f$ of the
quasiparticles is  determined  by the kinetic equation:
\begin{equation}
\dot f - {\partial \tilde E\over \partial {\bf r}} \cdot {\partial f\over
\partial
{\bf p}}+ {\partial \tilde E\over \partial {\bf p}} \cdot {\partial f\over
\partial
{\bf r}}={\cal J}_{coll}~.
\label{KineticEq}
\end{equation}
The collision integral conserves the momentum and the energy   of
quasiparticles,  i.e.
\begin{equation}
\sum_{\bf p}  {\bf p}  {\cal J}_{coll}=\sum_{\bf
p}  \tilde E ({\bf p}) {\cal J}_{coll}=\sum_{\bf
p}  E({\bf p})  {\cal J}_{coll}=0~,
\label{ConservationCollision}
\end{equation}
but not
necessarily the  number of quasiparticle: as a rule the quasiparticle
number is not
conserved in superfluids.

\subsection{Momentum exchange between superfluid vacuum and quasiparticles}

From the Eq.(\ref{ConservationCollision}) and from the two equations for
the superfluid
vacuum, Eqs.(\ref{ContinuityEquation},\ref{LondonEquation}),  one obtains
the time evolution
of the momentum density for each of two subsystems: the superfluid
background (vacuum) and
quasiparticles (matter). The momentum evolution of the superfluid vacuum is
\begin{equation}
m\partial_t (n {\bf v}_{{\rm s}})  =- m\nabla_i(J_i{\bf v}_{\rm s}) -n
 \nabla\left( {\partial \epsilon\over \partial n}  +\sum_{\bf p} f
{\partial E \over \partial n} \right) + P_i  \nabla  v_{{\rm s}i}~.
\label{SuperfluidMomentumEq}
\end{equation}
where ${\bf P}=m{\bf J}_{\rm q}$ is the momentum of lquid carried by
quasiparticles (see
Eq.(\ref{TotalCurrent2})),
while the evolution
of the  momentum density of quasiparticles:
\begin{equation}
\partial_t  {\bf P}   =\sum_{\bf p} {\bf p}  \partial_t f   =  -
\nabla_i(v_{{\rm s}i}{\bf P})    -\nabla_i \left(  \sum_{\bf p} {\bf p}f
{\partial   E\over
\partial p_i}\right)  -\sum_{\bf p}f
 \nabla   E  - P_i
 \nabla  v_{{\rm s}i}~.
\label{QuasiparticleMomentumEq}
\end{equation}

Though the momentum of each subsystem is not conserved because of the
interaction with the other subsystem, the  total momentum density of the
system,
superfluid vacuum + quasiparticles, is conserved:
\begin{equation}
m\partial_t J_i   = \partial_t (mn  v_{{\rm s}i} +P_i)=
-\nabla_i \Pi_{ik}~,
\label{TotalMomentumEq}
\end{equation}
with the stress tensor
\begin{equation}
\Pi_{ik} = m J_i v_{{\rm s}k} + v_{{\rm s}i} P_k+  \sum_{\bf p} p_k f
{\partial E\over
\partial p_i}  + \delta_{ik} \left(n\left({\partial \epsilon\over \partial
n}+ \sum_{\bf p} f{\partial E\over \partial n}\right) -
\epsilon\right)  ~.
\label{StressTensor}
\end{equation}

\subsection{Covariance vs conservation.}

The same happens with the energy. Energy and momentum can be exchanged between
the two subsystems of quasiparticles and superfluid vacuum in a way similar
to the
exchange of energy and momentum between matter and the gravitational field.
In the low energy limit, when the quasiparticles are ``relativistic'',
this exchange must be described in the general relativistic covariant form. The
Eq.(\ref{QuasiparticleMomentumEq}) for the  momentum density of
quasiparticles as well as the
corresponding equation for the quasiparticle energy density can be
represented as
\begin{equation}
T^\mu{}_{\nu;\mu}=0 ~~,~~{\rm or}~~ {1\over
\sqrt{-g}}\partial_\mu \left(T^\mu{}_\nu \sqrt{-g}\right) - {1\over
2}T^{\alpha\beta} \partial_\nu g_{\alpha\beta}= 0
 \,.
\label{CovariantConservation}
\end{equation}
This result does not depend on the
dynamics of the superfluid condensate (gravity field), which is not
``relativistic''. The
Eq.(\ref{CovariantConservation}) follows solely from the ``relativistic''
spectrum of
quasiparticles.

The Eq.(\ref{CovariantConservation}) does not represent any conservation in
a strict sense, since the covariant derivative is not a total derivative.
The extra term,
which is not the total derivative, describes the force acting on
quasiparticles
(matter) from the superfluid condensate (an effective gravitational field).
Since the
dynamics of the superfluid background is not covariant, it is impossible to
find such total
energy momentum tensor, $T^\mu{}_\nu({\rm total}) = T^\mu{}_\nu({\rm
quasiparticles})
+T^\mu{}_\nu({\rm background})$, which could have a covariant form and
simultaneously
satisfy the real conservation law $\partial_\mu T^\mu{}_\nu({\rm
total})=0$. The total
stress tensor in Eq.(\ref{StressTensor}) is evidently noncovariant.

But this is impossible even in the fully covariant Einstein gravity, where
one has an energy
momentum pseudotensor for the gravitational background. Probably this is an
indication that
the  Einstein gravity is really an effective theory. As we mentioned above,
effective
theories in condensed matter are full of such  contradictions related to
incompatible
symmetries. In a given case the general covariance is incompatible with the
conservation law;
in  cases of the vacuum energy (Sec.\ref{VacuumEnergyAnd}) and the Einstein
action
(Sec.\ref{EinsteinActionAnd}) the general covariance is incompatible with
the scale
invariance; in the case of an axial anomaly, which is also reproduced in
condensed matter
(see e.g.\cite{LammiTalk}), the conservation of the baryonic charge is
incompatible with
quantum mechanics; the  action of the Wess-Zumino type, which cannot be
written in 3+1
dimension in the covariant form (as we discussed at the end of
Sec.\ref{DynamicsSuperfluidVacuum}, Eq.(\ref{WessZumino})), is almost
typical phenomenon in
various condensed matter systems; the momentum density determined as
variation of the
hydrodynamic energy over
${\bf v}_{\rm s}$ does not coincide with the canonical momentum in many
condensed matter
systems; etc., there are many other examples of apparent inconsistencies in
the effective
theories of condensed matter. All such paradoxes arise due to reduction of
the degrees of
freedom in effective theory, and  they disappear completely (together with
some symmetries of
the low-energy physics) on the fundamental level, i.e. in a fully
microscopic description,
where all degrees of freedom are taken into account.

\subsection{Energy-momentum   tensor for ``matter''.}

Let us specify the tensor $T^\mu{}_\nu$ which enters
Eq.(\ref{CovariantConservation}) for
the simplest case, when the gravity is simulated by the superflow only, i.e. we
neglect the space-time dependence of the density
$n$ and of the speed of sound $c$. Then the constant factor $mnc$ can be
removed from the
metric in
Eqs.(\ref{ContravarianAcousticMetric}-\ref{CovarianAcousticMetric}) and the
effective metric is simplified:
\begin{equation}
 g^{00}=-1 ~,~ g^{0i}=-v_{\rm s}^i  ~,~ g^{ij}=
 c^2\delta^{ij} -v_{\rm s}^i v_{\rm s}^j  ~,
\label{ContravarianAcousticMetricReduced}
\end{equation}
\begin{equation}
 g_{00}=-\left(1-{{\bf v}_{\rm s}^2\over c^2}\right) ~,~
g_{0i}=-{v_{{\rm s} i}\over c^2} ~,~ g_{ij}= {1\over  c^2}\delta_{ij} ~,~
\sqrt{-g}={1\over c^3}~.
\label{CovarianAcousticMetricReduced}
\end{equation}
Then the energy-momentum tensor of quasiparticles can be represented as
\cite{FischerVolovik}
\begin{equation}
\sqrt{-g} T^\mu{}_\nu=\sum_{\bf p} f v_g^\mu p_\nu\,,\qquad
v_g^\mu v_{g\mu} = -1 +\frac1{c^2}
\frac{\partial E }{\partial p_i}
\frac{\partial E }{\partial p_i}\,,
\end{equation}
where  $p_0= - \tilde E$, $p^0= E$; the group four velocity is defined as
\begin{eqnarray}
v_g^i& =& \frac{\partial \tilde E}{\partial p_i}\,,\quad v_g^0=1\,,\quad
v_{gi}= \frac1{c^2} \frac{\partial  E }{\partial p_i}\,,
\quad v_{g0}=-\left(
1+\frac{{v_{\rm s}^i }}{c^2}  \frac{\partial E }{\partial
p_i}\right)\,.
\end{eqnarray}
Space-time  indices are throughout
assumed to be raised and lowered by the   metric in
Eqs.(\ref{ContravarianAcousticMetricReduced}-\ref{CovarianAcousticMetricReduced}
). The group
four velocity is null in the relativistic domain of the spectrum only:
$v_g^\mu v_{g\mu}=0$ if
$E =cp$.
The relevant components of the energy-momentum tensor are:
\begin{eqnarray}
\sqrt{-g} T^0{}_i & = & \sum_{\bf p} f {p}_i
= P_i\qquad \mbox{\sf momentum density in either frame},
 \nonumber\\
\quad -\sqrt{-g} T^0{}_0& = & \sum_{\bf p} f \tilde E\qquad
\mbox{\sf   energy density in laboratory frame},
\nonumber\\
\sqrt{-g} T^k{}_i & = &  \sum_{\bf p} f p_i v_g^k \qquad
\mbox{\sf   momentum flux in laboratory frame}, \nonumber\\
-\sqrt{-g} T^i{}_0 & = & -\sum_{\bf p} f \tilde E\frac{\partial E}{\partial
p_i}
= \sum_{\bf p} f \tilde E v_g^i\qquad  \mbox{\sf
energy flux in laboratory frame}, \nonumber\\
 \sqrt{-g} T^{00} & = & \sum_{\bf p} f  p^0 = \sum_{\bf p} f  E
 \qquad  \mbox{\sf  energy density in comoving frame}.
\end{eqnarray}
With this definition of the momentum-energy tensor the covariant
conservation law in
Eq.(\ref{CovariantConservation}) acquires the form:
\begin{equation}
(\sqrt{-g}T^\mu{}_\nu)_{,\mu}=\sum_{\bf p} f \partial_\nu \tilde E=  P_i
\partial_\nu v_{\rm
s}^i +
\sum_{\bf p} f |{\bf p}|\partial_\nu c
 \,.
\end{equation}
The right-hand side represents ``gravitational''   forces acting on the
``matter'' from the
superfluid vacuum.

\subsection{Local thermodynamic
equilibrium.}

Local thermodynamic equilibrium is characterized by the local
temperature $T$ and local normal component velocity ${\bf v}_{\rm n}$  in
Eq.(\ref{Equilibrium}).  In local thermodynamic equilibrium the components
of energy-momentum
for the quasiparticle system (matter) are determined by the generic
thermodynamic potential (the pressure), which has the form
\begin{equation}
\Omega=\mp T\frac1{(2\pi\hbar)^3} \sum_{s}  \int d^3p~{\rm ln}(1\mp f)\,,
\label{Pressure}
\end{equation}
with the upper sign for fermions and lower sign for bosons.
For phonons one has
\begin{equation}
\Omega=  \frac{\pi^2}{30\hbar^3}  T_{\rm eff}^4\,,\qquad
T_{\rm eff} = \frac{T}{\sqrt{1-w^2}}~\,,
\label{EffectiveT}
\end{equation}
where the renormalized effective temperature $T_{\rm eff}$ absorbs all the
dependence on
two velocities of liquid. The components of the energy momentum tensor are
given as
\begin{equation}
T^{\mu\nu}  =    (\varepsilon  + \Omega )u^\mu
u^\nu+\Omega g^{\mu\nu}\,,\qquad \varepsilon=-\Omega +T{\partial \Omega
\over\partial T}=3\Omega\,,\qquad T^\mu{}_\mu=0~.
\label{QuasipStressTensorRel2}
\end{equation}
where the four velocity of the ``matter'', $u^\alpha$ and
$u_\alpha=g_{\alpha\beta}u^{\beta}$, which satisfies the normalization
equation $u_\alpha u^\alpha=-1$, is expressed in terms of superfluid and
normal component
velocities as
\begin{equation}
u^0={1\over \sqrt{ 1 - w^2}}\,,\qquad u^i={v_{(n)}^i\over \sqrt{  1 -
w^2}}\,,\qquad u_i= { w_i \over \sqrt{{ 1 - w^2}}}\,,\qquad
u_0=-{1+{{\bf w} \cdot{\bf v}_{{\rm s}}} \over
\sqrt{ { 1 - w^2}}}\,.
\label{4Velocity}
\end{equation}

\subsection{Global thermodynamic
equilibrium.  Tolman temperature.}

The distribution of quasiparticles in local equilibrium in
Eq.(\ref{Equilibrium}) can be expressed via the
temperature four-vector $\beta^\mu $ and thus via the effective temperature
$T_{\rm eff}$:
\begin{equation}
f_{\cal T} = {1\over
1+\exp[-\beta^\mu p_\mu]}\,,\qquad \beta^\mu ={u^\mu\over  T_{\rm eff}}
=\left({1\over T}, {{\bf v}_n \over  T
}\right)\,,\qquad \beta^\mu\beta_\mu=-T_{\rm eff}^{-2}~.
\label{4Temperature}
\end{equation}

For the relativistic system, the true equilibrium with vanishing entropy
production is established if $\beta^\mu$ is a timelike Killing vector:
\begin{equation}
\beta_{\mu;\nu}+
\beta_{\nu;\mu}=0~,\;\;{\rm or}\qquad
\beta^\alpha\partial_\alpha g_{\mu\nu}+
(g_{\mu\alpha}\partial_\nu +g_{\nu\alpha}\partial_\mu)\beta^\alpha=0
~.
\label{EquilibriumConditions}
\end{equation}
For a time independent, space dependent situation the  condition
$0=\beta_{0;0}=\beta^i\partial_i g_{00}$ gives
$\beta^i=0$, while the other conditions are satisfied when
$\beta^0={\rm constant}$. Hence true equilibrium requires that
${\bf v}_{\rm n}=0$ in the frame where the superfluid velocity field is
time independent
(i.e. in the frame where $\partial_t{\bf v}_{\rm s}=0$), and
$T={\rm constant}$. These are just the global equilibrium conditions in
superfluids, at which no dissipation occurs.  From the   equilibrium
conditions $T={\rm
constant}$ and ${\bf v}_{\rm n}=0$ it follows that under the global
equilibrium the effective
temperature in Eqs.(\ref{EffectiveT}) is space dependent according to
\begin{equation}
T_{\rm eff}={T\over \sqrt{ 1 - v_{\rm s}^2}}={T\over \sqrt{-g_{00}}}\,.
\label{TolmanLaw}
\end{equation}
According to Eq.(\ref{4Temperature})  the
effective temperature $T_{\rm eff}$ corresponds to the ``covariant
relativistic'' temperature in general relativity. It is an apparent
temperature as measured
by the local observer, who ``lives'' in superfluid vacuum and uses sound
for communication as
we use the light signals. The Eq.(\ref{TolmanLaw}) corresponds to Tolman's
law in general
relativity. Note that in condensed matter the Tolman temperature is the
real temperature $T$
of the liquid.

\section{Horizons, ergoregions,  degenerate metric, vacuum instability and
all that.}

\subsection{Landau critical velocity,  event horizon and ergoregion}

If the superfluid velocity exceeds the Landau critical value
\begin{equation}
v_L={\rm min}{E({\bf p})\over p}
\label{LandauVelocity1}
\end{equation}
the energy $\tilde E({\bf p})$ of some excitations, as  measured in the
laboratory frame,  becomes negative. This allows for excitations to be
nucleated from the vacuum. For a superfluid velocity field which is
time-independent in the
laboratory frame, the surface $v_{\rm s}({\bf r})=v_L$, which bounds the
region where
quasiparticles can have negative energy, the ergoregion, is called the
ergosurface.

The behavior of the system depends crucially on the dispersion of the
spectrum at higher
energy. There  are two possible cases. The spectrum bents
upwards at high energy, i.e. $ E({\bf p})=cp +\gamma p^3$ with $\gamma >0$.
Such dispersion
is realized for the fermionic  quasiparticles in
$^3$He-A. They are ``relativistic'' in the low energy corner but become
``superluminal'' at
higher energy\cite{grishated}. In this case the Landau critical velocity
coincides with the ``speed of light'', $v_L=c$, so that the ergosurface is
determined by
$v_{\rm s}({\bf r})=c$.   In the Lorentz invariant limit of the energy much
below the
``Planck'' scale, i.e. at $p^2
\ll \gamma/c$, this corresponds  to the ergosurface at $g_{00}({\bf r})=0$,
which is just the
definition of the ergosurface in gravity.  In case of radial flow  of the
superfluid vacuum towards the origin, the ergosurface also represents the
horizon
in the Lorentz invariant limit, and the region inside the horizon
simulates a black hole for low energy phonons. Strictly
speaking this is not a true horizon for phonons: Due to the nonlinear
dispersion, their group velocity $v_g=dE/d p
=c+3\gamma p^2>c$, and thus the high energy quasiparticles are allowed to
leave the black hole region. It is, hence, a horizon only for quasiparticles
living exclusively in the very low energy corner: they are not aware
of the possibility of ``superluminal'' motion. Nevertheless, the mere
possibility to exchange
the information across the horizon allows us to construct the thermal state
on both sides of
the horizon (see Sec.\ref{ModTolmanLaw} below) and to investigate its
thermodynamics,
including the entropy related to the horizon \cite{FischerVolovik}.

In superfluid $^4$He the negative dispersion is realized, with the group
velocity
$v_g=dE/d p <c$. In such superfluids the ``relativistic'' ergosurface
$v_{\rm s}({\bf r})=c$ does not coincide with the true ergosurface, which is
determined by $v_{\rm s}({\bf r})=v_L< c$.  In superfluid $^4$He, the Landau
velocity is related to the roton part of the spectrum, and is about four
times less than $c$. In case of radial flow inward, the ergosphere
occurs at $v_{\rm s}(r)=v_L<c$, while the inner surface $v_{\rm s}(r)=c$
still marks the horizon. This is in contrast to relativistically invariant
systems,
for which ergosurface and horizon coincide for purely radial gravitational
field.
The surface $v_{\rm s}(r)=c$ stays a horizon even for excitations with very
high
momenta up to some critical value, at which the group velocity of
quasiparticle  again
approaches $c$.

\subsection{Painlev\'e-Gullstrand metric in effective gravity in superfluids.}

Let us consider the spherically symmetric radial flow of the superfluid
vacuum, which is
time-independent in the laboratory frame. Then dynamics of the phonon,
 propagating in this velocity field, is given by the line element provided
by the effective
metric in Eq.(\ref{CovarianAcousticMetricReduced}):
\begin{equation}
ds^2= -\left(1-{v_{\rm s}^2(r)\over c^2}\right)dt^2 + 2{v_{\rm s}(r)\over
c^2}drdt
+ {1\over c^2} (dr^2 + r^2 d\Omega^2 )\,.
\label{PaileveInterval}
\end{equation}

This equation corresponds to the Painlev\'e-Gullstrand line elements.  It
describes a black
hole horizon if the superflow is inward (see refs.\cite{unruh,vissersonic},
on the pedagogical review of Panlev\'e-Gullstrand metric see
\cite{Martel}). If $v_{\rm
s}(r)=-c(r_S/r)^{1/2}$ the flow simulates the  black hole in general
relativity. For the
outward superflow with, say,
$v_{\rm s}(r)=+c(r_S/r)^{1/2}$ the  white hole is reproduced. For the general
radial dependence of the superfluid velocity, the Schwarzschild radius
$r_S$ is determined as
$v_{\rm s}(r_S)=\pm c$; the  ``surface gravity''  at the Schwarzschild radius
is $\kappa_S  =(1/2)dv^2_{\rm s}/dr|_{r_S}$;  and the Hawking temperature
$T_{\rm H}=\hbar\kappa_S/2\pi$.

\subsection{Vacuum resistance to formation of horizon.}

It is not easy to create the flow with the horizon in the Bose liquid
because of the
hydrodynamic instability which takes place behind the horizon (see
\cite{Liberati}).  From Eqs.(\ref{ContinuityEquation}) and
(\ref{LondonEquation}) of
superfluid hydrodynamics at $T=0$ (which correspond to conventional
hydrodynamics of ideal
curl-free liquid) it follows that for stationary motion of the liquid one
has the relation
between $n$ and $v_{\rm s}$ along the streamline  \cite{FluidMechanics}:
\begin{equation}
{\partial(nv_{\rm s})\over \partial v_{\rm s}}=n\left(1-{v_{\rm s}^2\over
c^2}\right) \,.
\label{nvVSv}
\end{equation}
The current $J= nv_{\rm s}$ has a maximal value just at the horizon and
thus it must decrease behind the horizon, where $1-(v_{\rm s}^2/c^2)$ is
negative. This is,
however, impossible in the radial flow since, according to the continuity
equation
(\ref{ContinuityEquation}), one has $nv_{\rm s}=Const/r^2$ and thus the
current must
monotonically increase across the horizon. This marks  the hydrodynamic
instability behind
the horizon and shows that it is impossible to construct the
time-independent flow with the
horizon without the fine-tuning of an external force acting on the liquid
\cite{Liberati}.
Thus the liquid itself resists to the formation of the horizon.

Would the quantum vacuum always resist to formation of the horizon?
Fortunately, not. In
the considered case of superfluid $^4$He, the same ``speed of light'' $c$,
which  describes
the quasipartilces (acoustic waves) and thus determines the value of the
superfluid velocity
at horizon, also enters the hydrodynamic equations that establish the flow
pattern of the
``black hole''. In $^3$He-A these two speeds are well separated. The
``speed of light'' $c$
for quasiparticles, which determines the velocity of liquid flow at the
horizon, is  much less
than the speed of sound, which determines the hydrodynamic instabilities of
the liquid. That
is why there are no severe hydrodynamic constraints on the flow pattern,
the hydrodynamic
instability is never reached and the surface gravity at such horizons is
always finite.

However, even in such superfluids another instability can develop due to the
presence of a horizon \cite{KopninVolovik1998}. Usually the ``speed of
light'' $c$ for
``relativistic '' quasiparticles coincides with the critical velocity, at
which the
superfluid state of the liquid becomes unstable towards  the normal state
of the liquid. When
the superfluid velocity with respect to the normal component or
to the container walls exceeds $c$, the slope
$\partial J/ \partial v_{\rm s}$ becomes negative the superflow is locally
unstable.

Such superfluid instability, however, can be avoided if the container walls
are properly
isolated \cite{SimulationPainleve}. Then the reference frame imposed by the
container walls
is lost and the ``inner'' observer living in the superfluid does not know
that the superfluid
exceeded the Landau velocity and thus the treshold of instability. Formally
this means that
the superfluid  instability is regulated not by the superfluid velocity
field $v_{\rm s}$
(as in the case of the hydrodynamic instability discussed above) , but by
the velocity
${\bf w}={\bf v}_{\rm n}-{\bf v}_{\rm s}$ of the  counterflow between the
normal and
superfluid subsystems. A stable superfluid vacuum  can be determined as the
limit
$T\rightarrow 0$ at fixed ``subluminal'' counterflow velocity
$w<c$, even if the superfluid velocity itself is ``superluminal''. This can
be applied
also to quasiequilibrium vacuum state across the horizon, which is locally
the vacuum state as
observed by comoving ``inner'' observer. The superfluid motion in this
state is locally
stable, though slowly decelerates due to the quantum friction caused by
Hawking radiation and
other processes related to the horizon and ergoregion \cite{grishated}.

\subsection{Modified Tolman's law across the horizon.}\label{ModTolmanLaw}

The realization of the quasiequilibrium state across the horizon at nonzero
$T$  can be found
in Ref.\cite{FischerVolovik} for 1+1 case. In this state the superfluid
velocity is
``superluminal'' behind the horizon, $v_{\rm s}>c$, but the counterflow
is everywhere ``subluminal'': the counterflow velocity $w$ reaches maximim
value $w=c$ at the
horizon with $w<c$ both outside and inside the horizon.  The local
equilibrium with
the effective temperature $T_{\rm eff}$ in Eq.(\ref{EffectiveT}) is thus
determined on both
sides of the horizon. It is interesting that one has the  modified form of
the Tolman's law, which is valid on both sides of the horizon:
\begin{equation}
T_{\rm eff}={T_\infty\over \sqrt{ \big|1 - {v_{\rm s}^2\over
c^2}\big|}}={T_\infty\over
\sqrt{|g_{00}|}}\,.
\label{ModifiedTolmanLaw}
\end{equation}
Here $T_\infty$ is the temperature at infinity. The effective temperature
$T_{\rm eff}$, which
determines the local ``relativistic''  thermodynamics, becomes infinite at
the horizon, with
the cutoff determined by the nonlinear dispersion of the quasiparticle
spectrum at high
energy,
$\gamma>0$. The real  temperature $T$ of the liquid is continuous across
the horizon:
\begin{equation}
T= T_\infty~~~~{\rm at}~~v_{\rm s}^2< c^2~~~~~,~~~~ T= T_\infty  {c\over
|v_{\rm s}|}~~~~{\rm at}~~v_{\rm s}^2> c^2 \,.
\label{ModifiedTolmanLaw2}
\end{equation}

\subsection{One more vacuum instability: Painlev\'e-Gullstrand vs Schwarzschild
metric in effective gravity.}

In the effective theory of gravity, which occurs in condensed matter systems,
the   primary quantity is the contravariant metric tensor
$g^{\mu\nu}$ describing the energy spectrum. Due to this the
two seemingly equivalent representations of the black hole metric,
in terms of either the Schwarzschild or the
Painlev\'e-Gullstrand line elements, are in fact not
equivalent in terms of the required stability of the underlying
superfluid vacuum.

An ``equivalent'' representation of the black or white hole metric is given
by the Schwarzschild line element, which in terms of the same superfluid
velocity reads
\begin{equation}
 ds^2=-\left(1- v_{\rm s}^2/c^2\right)d\tilde t^2+{dr^2\over c^2-
v_{\rm s}^2} +(r^2/c^2)\,d\Omega^2\,.
\label{Schwarzschild}
\end{equation}
 The Eqs.(\ref{Schwarzschild}) and (\ref{PaileveInterval}) are related by the
coordinate transformation. Let us for simplicity consider the abstract flow
with the
velocity exactly simulating the Schwarzschild metric, i.e. $v^2_{\rm
s}(r)=r_S/r$ and we put
$c=1$. Then the coordinate transformation is
\begin{equation}
 \tilde t(r,t)=t +   \left({2\over v_{\rm s}(r)} + {\rm ln}~ {1- v_{\rm
s}(r)\over
1+v_s(r)}\right) ~,~d\tilde t=dt +{v_{\rm s}\over 1-v_{\rm s}^2}dr.
\label{Transformation}
\end{equation}
What is the difference between the Schwarzschild and Painlev\'e-Gullstrand
space-times in
the effective gravity? The Painlev\'e-Gullstrand metric is determined in
the ``absolute''
Newton's space-time  $(t,{\bf r})$ of the laboratory frame, i.e. as is
measured by the
external experimentalist, who lives in the real world of the laboratory and
investigates the
dynamics of quasipartcles using the physical laws obeying the Galilean
invariance of the
absolute space-time.  The effective Painlev\'e-Gullstrand metric, which
describes
the quasiparticle dynamics in the inhomogeneous liquid, originates from the
quasiparticle
spectrum
\begin{equation}
 E =v_{\rm s}(r)p_r \pm  cp~,
\label{SpectrumInSuperflow2}
\end{equation}
or
\begin{equation}
(E -v_{\rm s}(r)p_r)^2 =  c^2p^2~,
\label{SpectrumInSuperflow1}
\end{equation}
which determines the contravariant components of the metric. Thus the
energy spectrum in the
low-energy corner is the primary quantity, which determines the effective
metric for the
low-energy quasiparticles.

 The time $\tilde t$ in the Schwarzschild line element is the time as
measured by the ``inner''   observer at ``infinity''   (i.e. far from the
hole). The
``inner''  means that this observer `` lives''  in the superfluid
background and uses
``relativistic''   massless quasiparticles (phonons in $^4$He or
``relativistic'' fermionic
quasiparticles  in $^3$He-A) as a light for communication and for
sinchronization clocks. The
inner observer at some point
$R\gg 1$ sends  quasiparticles pulse at the moment
$t_1$ which arrives at point $r$ at
$t=t_1+\int_{r}^Rdr/|v_-|$ of the absolute (laboratory) time, where  $v_+$
and
$v_-$ are absolute (laboratory) velocities of radially propagating
quasiparticles,
moving outward and inward respectively
\begin{equation}
v_\pm ={dr\over dt}={dE\over dp_r}=\pm 1 +v_{\rm s}~.
\label{RadialVelocity}
\end{equation}
Since from the point of view of the inner observer the speed of light (i.e. the
speed of quasiparticles) is invariant quantity and   does not depend on
direction
of propagation, for him  the moment of arrival of pulse to $r$ is not $t$
but  $\tilde t =(t_1+t_2)/2$, where
$t_2$ is the time when the pulse reflected from $r$ returns to observer at
$R$. Since $t_2-t_1=\int_{r}^Rdr/|v_-|+\int_{r}^Rdr/|v_+|$, one obtains
for the time measured by inner observer as
\begin{equation}
 \tilde t(r,t)={t_1+t_2\over 2}=t +  {1\over 2} \left(\int^R_r{dr\over
v_+}+
\int^R_r{dr\over v_-}\right)=t +   \left({2\over v_{\rm s}(r)} + {\rm ln}~ {1-
v_s(r)\over 1+v_{\rm s}(r)}\right)-   \left({2\over v_{\rm s}(R)} + {\rm
ln}~ {1-
v_s(R)\over 1+v_{\rm s}(R)}\right)  ~,
\label{InnerTime}
\end{equation}
which is just the Eq.(\ref{Transformation}) up to a constant shift.

In the complete absolute physical space-time of the laboratory the external
observer
can detect quasiparticles  radially propagating into (but not out of) the
black hole  or out of (but not into) the white hole. The energy spectrum
of the quasiparticles remains to be  well determined both outside and
inside the horizon.  Quasiparticles cross the black hole horizon with the
absolute velocity
$v_-= -1-v_{\rm s}=-2$ i.e.   with the double speed of light: $r(t) = 1  -
2(t-t_0)$. In case of a white hole horizon one has $r(t) = 1  +
2(t-t_0)$.  On the contrary, from the point of view of the inner observer
the  horizon cannot be reached and crossed: the horizon can be approached
only asymptptically for infinite time: $r(\tilde
t)=1+(r_0-1)\exp(-\tilde t)$.  Such incompetence of the local observer, who
"lives" in the curved world of superfluid vacuum,  happens because he is
limited in his observations by the ``speed of light'', so that the
coordinate frame he uses is seriously crippled in the presence of the
horizon and becomes incomplete.

The Schwarzschild metric naturally arises for the inner observer, if the
Painlev\'e-Gullstrand metric is an effective metric for quasiparticles in
superfluids,  but not vice versa. The Schwarzschild metric
Eq.(\ref{Schwarzschild}) can in principle arise  as an effective metric in
absolute
space-time; however, in the presense of a horizon such metric indicates an
instability of the underlying medium. To obtain   a line element
of Schwarzschild metric as an effective metric for quasiparticles,  the
quasiparticle energy spectrum in the laboratory frame has to be
\begin{equation}
E^2=c^2\left(1-{r_S\over r}\right)^2 p_r^2+c^2\left(1-{r_S\over
r}\right) p_\perp^2~.
\label{SpectrumInSchwarzschild}
\end{equation}

In the presence of a horizon such spectrum has sections of the
transverse momentum $p_\perp$ with $E^2<0$. The imaginary
frequency of excitations
signals the instability of the superfluid vacuum if
this vacuum  exhibits the Schwarzschild metric as an effective metric for
excitations:  Quasiparticle
perturbations may
grow exponentially without bound in laboratory (Killing) time, as
$e^{t~{\rm Im} E}$, destroying the superfluid vacuum.  Nothing of this kind
happens in the case of the Painlev\'e-Gullstrand line element,
for which the quasiparticle energy is real even behind the horizon.
Thus the main difference between  Painlev\'e-Gullstrand and Schwarzschild
metrics as effective metrics is:  The first metric leads to the slow
process of the quasiparticle radiation from the vacuum at the horizon
(Hawking radiation), while the second one indicates a crucial instability
of the vacuum  behind the horizon.

In general relativity it is assumed that the two metrics can be converted
to each
other by the coordinate transformation in Eq.(\ref{Transformation}).
In condensed matter the coordinate transformation
leading from one metric to another is not that innocent
if an event horizon is
present.  The reason why the
physical behaviour implied by the choice of metric  representation
changes drastically is that the
transformation between the two line elements, $t\rightarrow t +\int^r
dr~v_{\rm s}/(c^2-v_{\rm s}^2)$, is singular on the horizon, and thus it
can be applied
only  to a part of the absolute space-time.  In condensed matter, only such
effective
metrics are physical which are determined everywhere in the real physical
space-time.  The two representations of the
``same'' metric cannot be strictly equivalent metrics, and we have
different classes of equivalence,
which cannot be transformed to each other by everywhere regular coordinate
transformation.  Painlev\'e-Gullstrand metrics for black and white holes
are determined everywhere, but belong to two different classes.  The
transition between these two metrics occurs  via the singular
transformation
$t\rightarrow t +2\int^r dr~v_{\rm s}/(c^2-v_{\rm s}^2)$ or via the
Schwarzschild line
element, which is prohibited in condensed matter physics, as explained
above,    since it is pathological in the presence of a horizon: it is not
determined in the whole space-time and it is singular at horizon.

\subsection{Incompleteness of space-time in effective gravity.}

It is also important that in the effective theory there is no need for the
additional extension of space-time to make it geodesically complete. The
effective space time
is always incomplete (open) in the presence of horizon, since it exists
only in the low energy
``relativistic'' corner and quasiparticles escape this space-time to a
nonrelativistic domain
when their energy increase beyond the relativistic linear approximation regime
\cite{grishated}.

Another example of the incomplete space-time in effective gravity is
provided by  vierbein
walls, or walls with the degenerate metric.  The physical origin of such
walls with the
degenerate metric  $g^{\mu\nu}$ in general relativity has been discussed by
Starobinsky at COSMION-99 \cite{Starobinsky}. They can arise after
inflation, if the inflaton
field has a $Z_2$ degenerate vacuum. The domain walls  separates the
domains with  2 diferent
vacua of the inflaton field. The metric $g^{\mu\nu}$ can everywhere satisfy
the Einstein equations in vacuum, but at the considered surfaces the
metric $g^{\mu\nu}$ cannot be diagonalized as
$g^{\mu\nu}={\rm diag}(-1,1,1,1)$. Instead, on such surface the metric is
diagonalized as $g^{\mu\nu}={\rm diag}(-1,0,1,1)$ and thus cannot be
inverted. Though
the space-time can be flat everywhere, the coordinate transformation cannot
remove
such a surface: it can only move the surface to infinity. Thus the system
of such
vierbein domain walls divides the space-time into domains which cannot
communicate with each other. Each domain is flat and infinite as viewed by
a local observer living in a given domain. In principle, the domains can have
different space-time topology, as is emphasized by Starobinsky
\cite{Starobinsky}.

In $^3$He-A such walls appear in a film of the $^3$He-A, which simulates the
2+1 vacuum. The wall is the topological solitons on which
one of the vectors (say, ${\bf e}_1$) of the order parameter playing the
part of the
vierbein in general relativity, changes sign across the wall
\cite{VierbeinWalls}:
\begin{equation}
{\bf e}_1(x)  =\hat{\bf x} c_0\tanh {x} ~,~{\bf e}_2 =\hat{\bf y} c_0~.
\label{VierbeinWall}
\end{equation}
The corresponding 2+1 effective metric experienced by quasiparticles, is
\begin{equation}
ds^2= -dt^2   +{1\over c_0^2} \left(dx^2 \tanh^{-2} {x}  +dy^2\right) ~.
\label{LineElement}
\end{equation}
Here $c_0$ is ``speed of light'' at infinity. The speed of ``light''
propagating
along the axis $x$  becomes zero at $x=0$, and thus $g_{xx}(x=0)=\infty$.
This indicates
that the low-energy quasiparticles cannot propagate across the wall.

The coordinate singularity at
$x=0$ cannot be removed by the coordinate transformation. If at
$x>0$ one introduces a new coordinate $\tilde x=\int dx/\tanh {x} $, then
the line element
acquires the standard  flat form
\begin{equation}
ds^2=-dt^2 + d\tilde x^2 +dy^2 ~.
\label{LineElementFlat}
\end{equation}
This means that for the ``inner'' observer, who measures the time and
distances using the
quasiparticles, his space-time is flat and infinite. But this is only half
of the real
(absolute)  space-time:  the other domain -- the left half-space
at $x<0$ -- which is  removed by the coordinate transformation, remains
completely unknown
to the observer living in the right half-space.   The situation is thus the
same as discussed
by Starobinsky for the domain wall in the inflaton field \cite{Starobinsky}.

Thus the vierbein wall divides the bulk liquid into two classically
separated flat
``worlds'', when viewed by the local ``inner''   observers who use the low
energy ``relativistic''   quasiparticles for communication. Such
quasiparticles cannot
cross the wall in the classical limit, so that the observers living on
different
sides of the wall cannot communicate with each other.  However, at the
``Planck scale'' the
quasiparticles have a superluminal dispersion in $^3$He-A, so that
quasiparticles with high
enough energy can cross the wall. This is an example of the situation, when
the effective space-time which is complete from the point of view of the
low energy observer
appears to be only a part of the more fundamental underlying space-time.
It is interesting
that when the chiral fermionic quasiparticles of $^3$He-A crosses the wall,
its chirality
changes to the opposite \cite{VierbeinWalls}: the lefthanded particle
viewed by the observer
in one world becomes the righthanded particle in the hidden neighbouring
world.

\section{Conclusion.}

We considered here only small part of the problems which arise in the
effective gravity of
superfluids. There are, for example, some other interesting effective
metrics, which must be
exploited. Quantized vortices with circulating superfluid velocity around
them simulate the
spinning cosmic strings, which experience the gravitational Aharonov-Bohm
effect measured in
superfluids as Iordanskii force acting on the vortex \cite{SpinningString};
the superfluid
vacuum around the rotating cylinder simulates
\cite{parallel} Zel'dovich-Starobinsky effect of radiation by the
dielectric object or black
hole rotating in quantum vacuum  \cite{Zeldovich2,Starobinskii}. The
expanding Bose
condensate in the laser manipulated traps, where the  speed of sound varyes
in time,  may
simulate the inflation. The practical realization of the analogue of event
horizon, the
observation of the Hawking radiation and measurement of the Bekenstein
entropy still remain a
challenge for the condensed matter physics (see Ref.\cite{JacobsonTalk} for
review of
different proposals).  However, even the theoretical consideration of the
effective gravity in
condensed matter can give insight into many unsolved problems in quantum
field theory. We can
expect that the analysis of the condensed matter analogues of the effective
gravity,  in
particular,  of the Landau-Khalatnikov two-fluid hydrodynamics
\cite{Khalatnikov} and its
extensions will allow us to solve the longstanding problem of the
cosmological constant.

\end{document}